%% file: main.tex
\documentclass[sigconf]{acmart}

\AtBeginDocument{%
  }

\copyrightyear{2026}
\acmYear{2026}
\setcopyright{cc}
\setcctype{by}
\acmConference[WWW '26]{Proceedings of the ACM Web Conference 2026}{April 13--17, 2026}{Dubai, United Arab Emirates}
\acmBooktitle{Proceedings of the ACM Web Conference 2026 (WWW '26), April 13--17, 2026, Dubai, United Arab Emirates}
\acmPrice{}
\acmDOI{10.1145/3774904.3793006}
\acmISBN{979-8-4007-2307-0/2026/04}

\definecolor{mycyan}{rgb}{0.0, 1.0, 1.0}     
\definecolor{myteal}{rgb}{0.0, 0.5, 0.5}

\usepackage{amsfonts}
\usepackage{mathtools}

\usepackage[inline]{enumitem}
\usepackage[skip=0pt]{caption}
\usepackage{multirow}
\usepackage{siunitx}
\usepackage{acronym}
\usepackage[x11names, rgb]{xcolor}
\usepackage{algorithm}
\usepackage{subcaption}

\usepackage{pgfplots}
\usepackage{tikz}
\usetikzlibrary{calc}
\pgfplotsset{compat=1.16}
\usepgfplotslibrary{groupplots}
\definecolor{sasrec}{RGB}{0,128,128}
\definecolor{random}{RGB}{255,182,193}
\definecolor{duor}{RGB}{0,130,128}

\pgfplotsset{
  sasrec/.style={
    fill={rgb,255:red,0;green,128;blue,128},
    draw=black,
    thick,
  },
  random/.style={
    fill={rgb,255:red,255;green,182;blue,193},
    draw=black,
    thick,
  },
  pct/.style={
    fill=purple!40!white,
    draw=black,
    thick,
  },
  pmmf/.style={
    fill=green!40!white,
    draw=black,
    thick,
  },
  ipr/.style={
    fill=yellow!40!white,
    draw=black,
    thick,
  }
  fair/.style={
    fill=orange!40!white,
    draw=black,
    thick,
  },
  duor/.style={
    fill=duor!40!white,
    draw=black,
    thick,
  },
  popsteer/.style={
    fill=blue!40!white,
    draw=black,
    thick,
  },
  legend rect/.style={
    legend image code/.code={
      \draw[fill=#1, draw=black, thick]
        (0cm,-0.07cm) rectangle (0.7cm,0.15cm); 
    }
  }
}

\usepackage{xspace}
\usepackage{booktabs,multirow,xcolor,siunitx}
\usepackage[table,x11names]{xcolor}
\definecolor{rowgray}{gray}{0.93}
\newcommand{\algname}[1] {{\fontfamily{cmtt}\selectfont {#1}}}

\begin{document}

\title[Interpretable Neuron Steering for Controlling Popularity Bias in Recommender Systems]{From Insight to Intervention: Interpretable Neuron Steering for Controlling Popularity Bias in Recommender Systems}

\author{Parviz Ahmadov}
\orcid{0009-0001-5334-1920}
\affiliation{%
  \institution{Delft University of Technology}
  \city{Delft}
  \country{The Netherlands}
}
\email{p.ahmadov@student.tudelft.nl}

\author{Masoud Mansoury}
\orcid{0000-0002-9938-0212}
\affiliation{%
  \institution{Delft University of Technology}
  \city{Delft}
  \country{The Netherlands}
}
\email{m.mansoury@tudelft.nl}

\begin{abstract}
  Popularity bias is a pervasive challenge in recommender systems, where a few popular items dominate attention while the majority of less popular items remain underexposed. This imbalance can reduce recommendation quality and lead to unfair item exposure. Although existing mitigation methods address this issue to some extent, they often lack transparency in how they operate. In this paper, we propose a post-hoc approach, \algname{PopSteer}, that leverages a Sparse Autoencoder (SAE) to both interpret and mitigate popularity bias in recommendation models. The SAE is trained to replicate a trained model’s behavior while enabling neuron-level interpretability. By introducing synthetic users with strong preferences for either popular or unpopular items, we identify neurons encoding popularity signals through their activation patterns. We then steer recommendations by adjusting the activations of the most biased neurons. Experiments on three public datasets with a sequential recommendation model demonstrate that \algname{PopSteer} significantly enhances fairness with minimal impact on accuracy, while providing interpretable insights and fine-grained control over the fairness–accuracy trade-off.
\end{abstract}



\maketitle

\section{Introduction}

Popularity bias is a persistent challenge in recommender systems, where highly popular items are favored over less popular, long-tail items that may be more relevant to niche user interests~\cite{canamares2018should, zhang2023empowering, liu2020long}. This bias arises naturally from collaborative filtering and other learning-based approaches that reflect patterns in historical interaction data, including skewness toward popular content~\cite{schnabel2016recommendations,huang2024going}. While such recommendations may serve mainstream users well, they limit opportunities for discovery, marginalize niche or newer items, and reduce exposure fairness. This can ultimately disadvantage both content creators and users with specialized tastes~\cite{zhu2022fighting,greenwood2024user}.

Most existing methods to mitigate popularity bias rely on reweighting item scores or modifying model architectures~\cite{zhu2021popularity,chen2021autodebias,prent2024correcting}. However, these techniques often function as black boxes, offering limited insight into why certain items are recommended or which internal components drive biased behavior. This lack of \textit{interpretability} poses a challenge for diagnosing and correcting systemic biases.

To address this, we turn to Sparse Autoencoders (SAEs)~\cite{ng2011sparse}—neural models designed to activate only a small subset of neurons per input. This sparsity enables clearer attribution of individual neurons to specific features or concepts. Recent work in interpretability has shown that SAEs can uncover high-level, monosemantic neurons in large language models~\cite{bricken2023monosemanticity, gao2024scaling, o2024disentangling, lan2024sparse}. Crucially, sparsity makes these neurons easier to isolate and manipulate without unintended side effects, a property already leveraged to mitigate bias in language models~\cite{durmus2024steering, hegde2024effectiveness, harle2024scar}.

Inspired by these findings, we propose \algname{PopSteer}, a novel post-hoc method that interprets and mitigates popularity bias at the neuron level in deep recommendation models. We begin by training an SAE to replicate the output of a pretrained recommender, attaching it to the final layer to capture the model’s decision-making process. Then, we generate synthetic user profiles that reflect strong preferences for either popular or unpopular items. By analyzing how individual SAE neurons respond to these inputs, we identify those most responsible for encoding popularity signals.

\begin{figure*}[t]
    \centering
    \includegraphics[width=.9\linewidth]{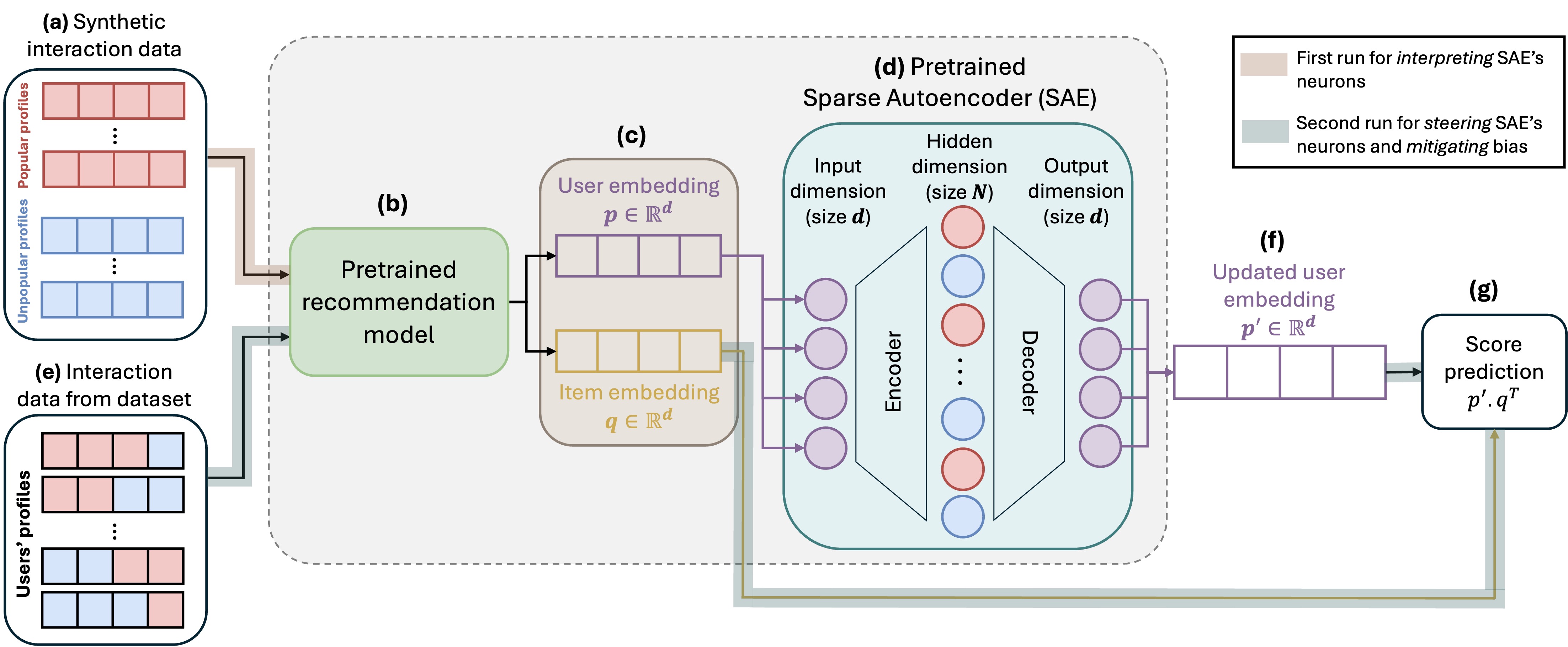}
    \vspace{3pt}
    \caption{Overview of \algname{PopSteer} process, first run with synthetic data for interpreting SAE's neuron ($a \rightarrow b \rightarrow c \rightarrow d$), and second run with real-world dataset for steering SAE's neurons to mitigate bias and derive modified user's embedding ($e \rightarrow b \rightarrow c \rightarrow d \rightarrow f \rightarrow g$): (a) synthetic interaction data containing users’ profiles extremely representing interest towards either popular or unpopular items, (b) a pretrained recommendation model trained on real (not synthetic) data, (c) users' and items' embeddings learned from the pretrained recommendation model, (d) SAE used for interpreting popularity bias according to activation behavior of neurons in hidden dimension, (e) users' profiles in real data, (f) updated user embedding after steering SAE's neurons, and (g) predicting the relevance score for a target user-item pair using the updated user embedding.}\label{fig:diagram}
\end{figure*}

Once identified, \algname{PopSteer} adjusts neuron activations to reduce the influence of popularity-biased neurons and amplify the contribution of neurons aligned with long-tail content. Figure \ref{fig:diagram} provides an overview of the \algname{PopSteer} process, illustrating both the interpretation and mitigation phases. This neuron steering approach enables fine-grained control over recommendation behavior while preserving the model's overall structure and performance. To the best of our knowledge, this is the first attempt for an interpretable solution to popularity bias in recommender systems. 

\algname{PopSteer} offers several advantages over existing popularity-bias mitigation methods:
(i) rather than reweighting interactions during training, it operates directly on learned embeddings, enabling popularity estimation on stable, post-trained representations;
(ii) it is transparent and interpretable, making it possible to examine the internal mechanism of the recommendation process and explicitly control its behavior;
(iii) it functions in a post-hoc manner, eliminating the need for model retraining and greatly improving practical deployability; and
(iv) unlike approaches that modify the training objective~\cite{ai2018unbiased,boratto2021connecting}, it requires no changes to the recommendation model’s loss function, enabling straightforward generalization to a wide range of model-based recommender systems.

Our experiments on three public datasets using the SASRec~\cite{kang2018self} model demonstrate that \algname{PopSteer} effectively improves exposure fairness for items with minimal loss in recommendation accuracy. Compared to several existing bias mitigation techniques, our method offers both superior performance and interpretability. 

\section{Sparse Autoencoder (SAE)}\label{sec:interpret_SAE}

Autoencoders are a class of neural networks designed to learn compressed representations of data~\cite{bank2023autoencoders,hinton2006reducing}. 
By training the network to minimize the reconstruction error, the autoencoder learns latent codes that capture the most salient features of the input, often revealing underlying structure in the data. While classical autoencoders can reveal structure in the input data, their hidden units often exhibit dense, overlapping activations, making the learned features difficult to interpret and sometimes redundant~\cite{laakom2024reducing}. 

To address this, \emph{Sparse Autoencoders} (SAEs), introduced by~\citet{ng2011sparse}, impose sparsity constraints on hidden unit activations, ensuring only a small subset of neurons fire for any given input. By enforcing sparsity, the model is encouraged to discover more discriminative and disentangled features, improving robustness and uncovering underlying structures in the data.

Formally, let $x \in \mathbb{R}^d$ denote an embedding from the machine learning (ML) model. The SAE, consisting of an input dimension of size $d$, a hidden dimension of size $N$, and an output dimension of size $d$, reconstructs $x$ as:
\begin{equation}\label{eq_sae_xhat}
\hat{x} = W_{\text{dec}} a + b_{\text{pre}} 
\end{equation}
\noindent where $W_{\text{dec}} \in \mathbb{R}^{d \times N}$ and $b_{\text{pre}} \in \mathbb{R}^d$ are learnable parameters. The sparse representation $a \in \mathbb{R}^N$ is computed as:
\begin{equation}\label{eq_sae_a} 
a = \text{TopAct}(z) 
\end{equation}
\begin{equation*} 
z =W^T_{\text{enc}} (x - b_{\text{pre}})
\end{equation*} 

\noindent where $W_{\text{enc}} \in \mathbb{R}^{d \times N}$ is the learnable encoder weight matrix and TopAct is the activation function retaining $K$ neurons with highest activation weights\footnote{This function is commonly referred to as TopK in the sparse autoencoder literature \cite{gao2024scaling}. We denote it as TopAct here to avoid confusion with the standard "top-$k$" terminology used for item recommendation in recommender systems.}. Here $z \in \mathbb{R}^N$ are the \emph{hidden activations}, where our neuron analysis and interventions occur, and $a$ is the \emph{sparse activations} after TopAct.
The sparsity is enforced through retaining only the $K$ highest activations in $z$ and setting all other activations to zero, as proposed by Gao et al.~\cite{gao2024scaling}. Choosing $K\ll N$, limits each input to at most $K$ active neurons while giving the model a large over-complete pool of candidates from which to “select”. 

This combination of over-completeness and strict sparsity pushes every firing unit to \textit{specialize} in a single, interpretable feature. Empirically, the resulting representations are largely monosemantic—individual neurons correlate with one concept and exhibit minimal dependence on their peers~\cite{bricken2023monosemanticity, gao2024scaling}. This allows for cleaner intereference through neuron steering, with minimal impact on performance. SAE is trained to minimize the following objective: 
\[
\begin{aligned}
\min_{W_{\text{enc}}, W_{\text{dec}}, b_{\text{pre}}}
    &\; \|x-\hat{x}\|_2^2 + \gamma\,\mathcal{L}_{\text{aux}}
\end{aligned}
\]
\noindent where $\mathcal{L}_{aux}$ is a loss term preventing dead neurons and $\gamma$ balances primary reconstruction loss and reviving dead neurons. A neuron is flagged as dead if it has not activated for a predetermined number of inputs (e.g., 10 million tokens), preventing it from receiving meaningful gradient updates. $\mathcal{L}_{aux}$ targets dead neurons specifically, as detailed in \cite{gao2024scaling}, to encourage them to learn useful features without affecting active ones and is defined as $\mathcal{L}_{\text{aux}} = \|e-\hat{e}\|_2^2$ where $e=x-\hat{x}$, $\hat{e}=W_{\text{dec}} z'$, and $z'=\text{TopAct}_{aux} \big( z \big)$.
This objective ensures that the sparse hidden representation maintains fidelity to the original embedding while revealing the underlying decision structure. The learned neurons can then be interpreted to explain which concepts in the input data influence the ML model's output.

\section{\algname{PopSteer} method}

In this section, we describe \algname{PopSteer}, our proposed method for interpreting and mitigating popularity bias. \algname{PopSteer} operates on a pre-trained recommendation model together with a Sparse Autoencoder (SAE). Given this pipeline, \algname{PopSteer} focuses on tracking and interpreting the hidden neurons of the SAE. Since each SAE neuron tends to specialize in a distinct concept derived from the input data, analyzing its activation behavior provides insights into how such concepts influence the model’s predictions.

To interpret popularity-related concepts captured by these neurons, we require user profiles that clearly reflect preferences toward either popular or unpopular items. For this purpose, we generate synthetic user profiles that explicitly highlight popularity bias and examine how specific neurons respond to them (as illustrated in the first phase of Fig.~\ref{fig:diagram}, steps  $a \rightarrow b \rightarrow c \rightarrow d$). The motivation for creating synthetic profiles is to provide controlled inputs with strong popularity-aligned and unpopularity-aligned signals, enabling clearer neuron-level interpretation and isolation of popularity-related features.

Each synthetic profile is passed through the pre-trained recommendation model to obtain a user embedding (i.e., $x$ in Section~\ref{sec:interpret_SAE}), which is then fed into the pre-trained SAE for neuron-level analysis. In the following, we describe our synthetic data generation process and how we quantify each neuron’s contribution to popularity bias.

\subsection{Generating synthetic profiles}\label{sec_synthetic}

We create two types of synthetic user profiles: one biased toward popular items and the other toward unpopular items. These profiles simulate user interactions that reflect extreme cases of popularity preference, enabling clearer observation of neuron activation.

Let $\mathcal{U} = \{u_1,\dots,u_n\}$ be the set of users, $\mathcal{I} = \{i_1,\dots,i_m\}$ the set of items, and $R \in \mathbb{R}^{n \times m}$ the user–item interaction matrix. Following~\cite{abdollahpouri2021user,zhang2023empowering}, we define the popular (\textit{head}) item set $\mathcal{I}^{\text{Pop}}$ as the top 10\% most frequently interacted items and the unpopular (\textit{tail}) item set $\mathcal{I}^{\text{Unpop}}$ as the bottom 10\%\footnote{These ratios are used in our experiments, but other thresholds are possible.}. Based on these sets, we construct two synthetic datasets: (1) $R^{\text{Pop}}$, where each user profile contains only popular items ($\mathcal{I}^{\text{Pop}}$), and (2) $R^{\text{Unpop}}$, where each profile contains only unpopular items ($\mathcal{I}^{\text{Unpop}}$).

To construct \(R^{\text{Pop}}\), we generate \(n^\prime\) synthetic users' profiles, each of length \(M\).
For each synthetic user, we randomly sample \(M\) items (with replacement) from \(\mathcal{I}^{\text{Pop}}\) (see Section~\ref{sec_expt_setup} for detailed experimental setup). 
The construction of \(R^{\text{Unpop}}\) is identical, except that items are drawn from the tail set \(\mathcal{I}^{\text{Unpop}}\). By restricting each synthetic profile to \emph{either} exclusively head items \(\mathcal{I}^{\text{Pop}}\) \emph{or} exclusively tail items \(\mathcal{I}^{\text{Unpop}}\), we create two synthetic sets that lie at the extremes of the popularity spectrum. 

This design cleanly isolates the influence of item popularity on model behaviour while suppressing confounding factors such as sequence diversity, recency effects, and mixed user interests. Importantly, these synthetic datasets are not used to train the SAE. Instead, we feed them into a pretrained SAE (trained on $R$), to evaluate its neurons' activation, ensuring that its learned representations remain grounded in real user behavior.

A natural concern is that such synthetic profiles may appear unrealistic compared to real-world usage patterns, where users interact with a mix of popular and less popular items. We confirm that our synthetic users are intentionally extreme. This extremization is by design: it amplifies the influence of popularity relative to other latent factors and makes it easier to separate the effect of popularity from other causes. Our goal here is not to model user behavior, but to produce a clear diagnostic signal that reveals which neurons respond most strongly to popularity.

\subsection{Detecting Neurons Encoding Popularity Bias}
Each synthetic dataset is fed into the pretrained SAE, and the activation levels of the hidden layer neurons are recorded. To quantify how much each SAE neuron contributes to popularity bias, we compare neuron activations when processing $R^{\text{Pop}}$ and $R^{\text{Unpop}}$. 

Prior work suggests that as neural networks grow in size, neuron activation distributions tend to approximate a Gaussian distribution~\cite{lee2017deep, wolinski2022gaussian, haider2025neurons}. To sanity-check this assumption for our SAE, we compute skewness and excess kurtosis for hidden activation $z_j$ for each neuron $j$ and summarise the results in Table~\ref{tab:sae-normality}. For each dataset we report the percentage of neurons with $|\text{skew}| < 0.5$ and $|\text{excess kurtosis}| < 1$, which are commonly used ranges for approximate normality \cite{haider2025neurons}. Across ML-1M, BeerAdvocate, and Yelp, between $78.65\%$ and $92.53\%$ of neurons satisfy the skewness criterion and between $92.37\%$ and $97.59\%$ satisfy the kurtosis criterion, so treating the activation distributions as approximately Gaussian is a reasonable approximation in our setting.

\begin{table}[t]
    \centering
    \setlength{\tabcolsep}{4pt} 
    \small                      
    \caption{Normality diagnostics for SAE neuron activations. Percentage of neurons whose skewness and excess kurtosis fall within typical ranges
    for approximate normality.}
    \label{tab:sae-normality}
    \begin{tabular}{lcc}
        \toprule
        Dataset      & \% with $|\text{skew}| < 0.5$ & \% with $|\text{excess kurtosis}| < 1$ \\
        \midrule
        ML-1M        & 87.28                         & 95.83                                  \\
        BeerAdvocate & 92.53                         & 97.59                                  \\
        Yelp     & 78.65                         & 92.37                                  \\
        \bottomrule
    \end{tabular}
\end{table}

We leverage this property to use Cohen’s $d$~\cite{cohen2013statistical} to measure the effect size of activation differences for each neuron $j$:
\begin{equation}\label{eq_d}
d_j = \frac{\mu_{j,\text{Pop}} - \mu_{j,\text{Unpop}}}{\sqrt{\frac{\sigma^2_{j,\text{Pop}} + \sigma^2_{j,\text{Unpop}}}{2}}} 
\end{equation}
\noindent where $\mu_{j,\text{Pop}}$ and $\mu_{j,\text{Unpop}}$ are the mean activations of neuron $j$ under $R^{\text{Pop}}$ and $R^{\text{Unpop}}$, respectively, and $\sigma_{j,\text{Pop}}$ and $\sigma_{j,\text{Unpop}}$ are the corresponding standard deviations.
Cohen’s $d$ provides a normalized measure of the difference between two distributions. We denote Cohen's $d$ value of neuron $j$ as $d_j$. A high absolute value of $d_j$ indicates that neuron $j$ is strongly responsive to popularity-related patterns. Positive values suggest alignment with popular content, while negative values imply a focus on unpopular or niche items.

\subsection{Steering Neurons} \label{popularity_mitigation}

To counteract popularity bias in recommendations, we use the computed Cohen's $d$ values to identify neurons most responsible for encoding popularity signals. Our method, \algname{PopSteer}, applies a process called \emph{neuron steering}, which systematically adjusts these neuron activations to reduce bias and promote fairer item exposure (refer to the second phase in Fig.~\ref{fig:diagram}: steps e→b→c→d→f→g for the steering process on real data).

Neuron steering modifies the hidden activations within the SAE to either amplify or suppress the influence of popularity-related neurons. The adjustment is guided by each neuron's $d_j$ score, which reflects its alignment with popular or unpopular items. We modify the original activation $z_j$ of neuron $j$ to obtain a new activation $z_j^\prime$: 
\begin{equation}\label{steering_eq}
z_j^\prime = 
\begin{cases} 
z_j - w_j \cdot \sigma_j, & \text{if } d_j > \beta \\
z_j + w_j \cdot \sigma_j, & \text{if } d_j < -\beta 
\end{cases} 
\end{equation}

\noindent where $\sigma_j$ is the standard deviation of activations for neuron $j$ and $\beta$ is a hyperparameter controlling the threshold on $|d_j|$ for selecting neurons to steer, ensuring adjustments are applied only to those with sufficiently strong bias signals ($\beta=0$ steers all neurons). $w_j$ is a weight assigned to each neuron $j$ based on the normalized absolute value of its $d_j$. This weight determines how much each neuron's activation will be adjusted:

\begin{equation}\label{eq_w}
w_j = 
\begin{cases} 
\alpha^{\text{Pop}} \cdot \frac{|d_j|}{\max\limits_{0 \leq i \leq N}(|d_i|)} & \text{if } d_j > \beta  \\
\alpha^{\text{Unpop}} \cdot \frac{|d_j|}{\max\limits_{0 \leq i \leq N}(|d_i|)} & \text{if } d_j < -\beta 
\end{cases}
\end{equation}

\noindent where $\alpha^{\text{Pop}}$ and $\alpha^{\text{Unpop}}$ are tunable hyperparameters that independently control the strength of the steering for popularity-aligned and unpopularity-aligned neurons, respectively.
We use different steering strengths for popularity-aligned and unpopularity-aligned neurons because we found that these two groups respond differently to the same steering magnitude, as further discussed in Section \ref{sensitivity_analysis}. In this formulation, neurons with stronger associations to popularity bias (higher $|d_j|$) receive larger weights ($w_j$).

This adjustment boosts neurons promoting unpopular items ($d_j < -\beta$) and suppresses those favoring popular items ($d_j >\beta$). 
After steering the hidden layer activations and applying TopAct function (Eq.~\ref{eq_sae_a}), the updated SAE then returns the modified user embedding $p'$ (part (f) in Fig.~\ref{fig:diagram}) according to Eq.~\ref{eq_sae_xhat}.
This steered embedding is used together with item embeddings from the base recommendation model to generate the final recommendation list (part (g) in Fig.~\ref{fig:diagram}).

Our steering strategy differs from existing methods in the following ways. First, many approaches either zero out the neuron activation or clamp it to a constant \cite{durmus2024steering, hegde2024effectiveness, kang2024interpret, zhao2024steering}. Such hard interventions can strongly distort the activation distribution and have undesirable side effects. We instead scale interventions by the standard deviation of each neuron, so that adjustments are proportional to the usual spread of its activations. 
Consistent with this design, Haider et al.~\cite{haider2025neurons} compare activation-based interventions to zeroing strategy in neural networks and report that the former results in fewer adverse effects. Second, beyond variance, we account for the amount of bias encoded by each neuron through Cohen's $d$ and use this effect size as its steering weight.

\section{Experiments}

Our experimental analysis on real-world datasets is designed to address the following research questions\footnote{Due to space constraints, scalability analysis of \algname{PopSteer} is presented in Appendix~\ref{scalability_analysis}.}: \textbf{(RQ1)} Does our \algname{PopSteer} method outperform baselines in improving fairness of exposure for items with negligible loss in ranking quality? \textbf{(RQ2)} How effective is our \algname{PopSteer} method in interpreting recommendation models with respect to popularity bias? \textbf{(RQ3)} How does varying the neuron adjustment hyperparameters influence the performance of our \algname{PopSteer} method?

\subsection{Datasets}
We conduct experiments on three public datasets from diverse domains: 
\begin{enumerate*}[label=(\roman*)]
    \item MovieLens 1M (ML-1M)~\cite{harper2015movielens}, a movie recommendation dataset,
    \item BeerAdvocate~\cite{McAuleyLeskovec2013Expertise}, a beer review dataset
    \item Yelp~\cite{yelp_open_dataset}, a restaurant recommendation dataset.
\end{enumerate*}

Following standard preprocessing~\cite{he2017translation, kang2018self}, we create a 5-core sample on ML-1M, and BeerAdvocate datasets, where each user has at least 5 interactions and each item is interacted with by at least 5 users. For Yelp dataset, we perform 20-core validation due to extreme sparsity. All datasets include timestamp information, enabling their use in sequential recommendation task. Table~\ref{table_datastats} summarizes the statistical properties of these datasets.

\begin{table}[t!]
  \centering
  \caption{Statistics of the datasets.}
  \begin{tabular}{lrrrr}
    \toprule
    Dataset & \#Users & \#Items & \#Interactions & Density \\
    \midrule
    ML-1M    & 6,040    & 3,417    & 999,611        & 4.8\% \\
    BeerAdvocate    & 10,464    & 13,907    & 1,395,865   & 1\% \\
    Yelp  & 20,799    & 16,253    & 983,530         & 0.29\% \\
    \bottomrule
    \label{table_datastats}
  \end{tabular}
\end{table}

\subsection{Baselines}\label{sec_baselines}

We compare our \algname{PopSteer} method with the following baselines:
\begin{itemize}
    \item \textbf{Provider Max-Min Fairness (\algname{P-MMF}) \cite{xu2023p}:} Max-min fairness re-ranking method that adds a penalty on the gap between each provider’s exposure and the current worst-off provider. We apply \algname{P-MMF} with Min-regularize heuristic in our experiment. We tune the tradeoff hyperparameter $\lambda \in \{0.01, 0.05, 0.1, 0.2, 0.4, 0.6, 0.8, 1.0\}$.
    \item \textbf{Personalized Calibration Targets (\algname{PCT})~\cite{wang2023two}:} A two-sided method that balances user-level and system-level exposure. A solver computes exposure targets using linear programming, followed by a reranker that modifies each user’s recommendation list using a modified MMR strategy. Hyperparameters are tuned similar to \algname{FA*IR}. 
    \item \textbf{Inverse Popularity Ranking (\algname{IPR}) \cite{zhang2010niche}:} A reweighting approach that adjusts relevance score for user-item pairs based on inverse popularity: $\tilde{s}_{u,i}= s_{u,i}/\!\left(1+\alpha\,\rho_i\right)$, where $\rho_i=\mathrm{pop}(i)\big/\max_{j\in\mathcal{I}}\mathrm{pop}(j)$ and $\alpha$ is a hyperparameter controlling the degree of mitigating bias. We tune $\alpha \in \{0.01, 0.05, 0.1,\allowbreak 0.2, 0.4, 0.6, 0.8, 1.0\}$.
    \item \textbf{Fair Top-k Ranking (\algname{FA*IR})~\cite{zehlike2017fa}:} A post-processing approach that enforces fairness by ensuring a minimum proportion of unpopular items appear within each prefix of the top-k list. In our experiment, we set the proportion of unpopular items to $p \in \{0.01, 0.05, 0.1, 0.2, 0.4, 0.6, 0.8, 1.0\}$ and the significance level to $\alpha \in \{0.01,0.05,0.1\}$.
    \item \textbf{Dynamic User-oriented re-Ranking (\algname{DUOR})~\cite{gulsoy2025duor}:} A post-processing re-ranking approach that dynamically calibrates recommendations to align with each user's popularity inclination. Items are categorized into popular and unpopular sets via the Pareto Principle, and the method iteratively adjusts the recommendation list by penalizing or promoting items based on the current list's popularity ratio compared to the user's popularity inclination. We tune \algname{DUOR} with long recommendation lists of size $t \in \{50, 100, 250, 500, 1000\}$
\end{itemize}

We also compare our \algname{PopSteer} with a simple \algname{Random} baseline that randomly selects $k$ items from an initial longer recommendation list of varying size $t \in \{15,30,50,75,100\}$. We set the size of long recommendation lists for \algname{P-MMF}, \algname{PCT}, \algname{IPR}, and \algname{FA*IR} to 250.

\subsection{Evaluation metrics}\label{sec_metrics}

We evaluate recommendation performance along two dimensions: ranking performance and fairness metrics. For measuring ranking quality, we use Normalized Cumulative Discount Gain (nDCG). We additionally measure nDCG separately on Head and Tail items to evaluate the debiasing effect of \algname{PopSteer} (see Appendix~\ref{sec_debiasing}).

To evaluate fairness, we consider the following well-known metics for measuring exposure fairness for items~\cite{antikacioglu2017post,mansoury2022understanding}: 
\begin{itemize}
    \item \textbf{Item Coverage:} The percentage of items that appeared in recommendation lists. For reliable evaluation, following~\cite{mansoury2021graph,mansoury2022understanding}, we measure this metrics as percentage of items that appeared at least 5 times in recommendation lists.
    \item \textbf{Gini Index:} The measure of distribution uniformity for recommended items. Given distribution of recommended items representing how many times each items is recommended, Gini Index measures how uniform this distribution is. Uniform distribution (Gini index equal to zero) signifies that all items are equally represented in recommendation lists which is the ideal case.
\end{itemize}

\subsection{Experimental setup}\label{sec_expt_setup}
We adopt the widely used leave-one-out evaluation protocol, holding out the most recent interaction for testing, the second most recent for validation, and using the rest for training \cite{kang2018self}. The final recommendation list size is set to 10 for all our experimental results. For synthetic data generation process described in Section~\ref{sec_synthetic}, we set the number of synthetic users $n^\prime=409,600$ (200 epochs with 2048 users in each epoch), and the size of user's profile $M=50$\footnote{$M$ is the largest number of recent items the base recommender can process at once.}. 

For our experiments, we use SASRec~\cite{kang2018self} as a base model, whose self-attention captures temporal dependencies in user interactions. We adopt a sequential backbone because it simplifies new-user creation: a synthetic profile is specified directly by an item sequence.  
To test the generalizability of \algname{PopSteer}, we also conducted experiments using LightGCN \cite{he2020lightgcn}, graph-based general recommender system as a base model. Rather than constructing synthetic user sequences, we labeled \emph{existing} users based on their recent interaction histories. Due to space constraints, the results of these experiments are presented in our Github repository\footnote{\href{https://github.com/parepic/PopSteer2.0}{https://github.com/parepic/PopSteer2.0}}.

SASRec is trained with a learning rate of 0.001 and early stopping (patience = 10 epochs) based on validation nDCG. Model hyperparameters of SASRec is provided in Appendix \ref{sasrec_details}. The SAE includes two key hyperparameters: the scale factor $s \in \{32, 64, 96, 128\}$, which determines the size of the hidden layer, and the sparsity level $K \in \{36, 40, 44, 48, 52, 56\}$, which specifies the number of top activations retained in the hidden layer. Both SASRec and SAE are trained with the Adam optimizer and a batch size of 2048.

\input{fig_baselines_gini_}

\algname{PopSteer} involves multiple hyperparameters. $\alpha^{\text{Pop}}$ and $\alpha^{\text{Unpop}}$ as defined in Eq.~\ref{eq_w} rescale the per-neuron weight $w_j$ for neurons with $d_j>\beta$ (aligned with popular items) and $d_j<-\beta$ (aligned with unpopular items), respectively. Larger $\alpha^{\text{Pop}}$ produces a stronger downward adjustment for popularity-aligned neurons; larger $\alpha^{\text{Unpop}}$ produces a stronger upward adjustment for unpopularity-aligned neurons. 
We tune $\alpha^{\text{Pop}}, \alpha^{\text{Unpop}} \in \{1.0,\allowbreak 1.5,\allowbreak 2.0,\allowbreak 2.5,\allowbreak 3.0,\allowbreak 3.5,\allowbreak 4.0,\allowbreak 4.5,\allowbreak 5.0\}$.
The third hyperparameter involved in \algname{PopSteer} is a threshold $\beta$ that determines which neurons are modified: we steer all neurons whose Cohen's $d$ exceeds the threshold in magnitude, i.e., $\{\,j : |d_j| \ge \beta\,\}$. Larger $\beta$ yields fewer steered neurons. We tune $\beta \in \{0, 0.5, 1.0, 1.5, 2.0\}$. \algname{PopSteer} and the baselines are tuned using grid-search. Our implementation and code are available at \href{https://github.com/parepic/PopSteer2.0}{https://github.com/parepic/PopSteer2.0}.

\section{Results}

In this section, we provide evidence and observations from our experimental results to address our research questions.

\subsection{SAE performance}\label{performance}
We assess how well the SAE preserves user information by measuring the cosine similarity between original and reconstructed user embeddings. We also evaluate downstream recommendation quality by comparing nDCG scores when the model uses original versus reconstructed embeddings. The results appear in Table~\ref{tab:sae-performance}.

\begin{table}[t!]
    \centering
    \caption{SAE reconstruction quality and nDCG (original vs.\ reconstructed embeddings).}
    \label{tab:sae-performance}
    \begin{tabular}{lccc}
        \toprule
        Dataset       & Cosine sim. & nDCG (orig.) & nDCG (recon.) \\
        \midrule
        ML-1M         & 0.998       & 0.1204           & 0.1202            \\
        BeerAdvocate  & 0.984       & 0.0334           & 0.0333            \\
        Yelp      & 0.989       & 0.0202           & 0.0197            \\
        \bottomrule
    \end{tabular}
\end{table}

Across datasets, the mean cosine similarity between original and reconstructed user embeddings remains above 0.98, validating that the underlying user representation is preserved. Furthermore, the negligible drop in nDCG supports the conclusion that the downstream impact on performance is minimal.

\subsection{Overall performance}\label{performance}

In this section, we answer \textbf{RQ1}. Fig.~\ref{fig:baseline_tradeoffs} present our main results, comparing \algname{PopSteer} with SASRec and several baselines across three datasets in terms of nDCG versus Item Coverage (Fig.~\ref{fig:baseline_cov}) and nDCG versus Gini Index (Fig.~\ref{fig:baseline_gini}). 
Each point represents a hyperparameter configuration, with superior methods achieve points that balance high accuracy and fairness. The best performance in Figs~\ref{fig:baseline_cov} and~\ref{fig:baseline_gini} appear in the upper-right and bottom-right, respectively.

On ML-1M and BeerAdvocate, \algname{PopSteer} consistently outperforms all baselines in enhancing fairness while maintaining recommendation accuracy. It is evident that with almost the same nDCG, \algname{PopSteer} yields significantly higher Item Coverage and lower Gini Index compared to the baselines.

On Yelp dataset, while \algname{IPR} and \algname{DUOR} achieve higher fairness gains compared to \algname{PopSteer}, this improvement is at the cost of higher drop in nDCG. In a more realistic and lower nDCG loss situation, \algname{PopSteer} is still superior, outperforming all baselines in terms of both Item Coverage and Gini Index. Besides fairness improvements, \algname{PopSteer} also increases ranking quality of SASRec by $\sim$2.5\%, indicative of its strong effectiveness in improving recommendation overall performance (both accuracy and fairness). 

Another notable pattern is the stability of \algname{PopSteer}'s trade-off curves across datasets. While baselines often show sharp accuracy drops for only slight fairness gains, \algname{PopSteer} maintains gradual improvements. This stability shows the reliability of adjustments on targeted neuron, which minimize unintended disruptions to the model's core decision-making process compared to other approaches. This controllability is advantageous for deployment, as it allows practitioners to easily and reliably tune the trade-off between fairness and accuracy according to specific application needs without unpredictable outcomes. Beyond fairness gains, Appendix~\ref{sec_debiasing} presents experimental results for debiasing effect of \algname{PopSteer}. 

\begin{figure}[t]
    \centering
    \includegraphics[width=\linewidth
    ]{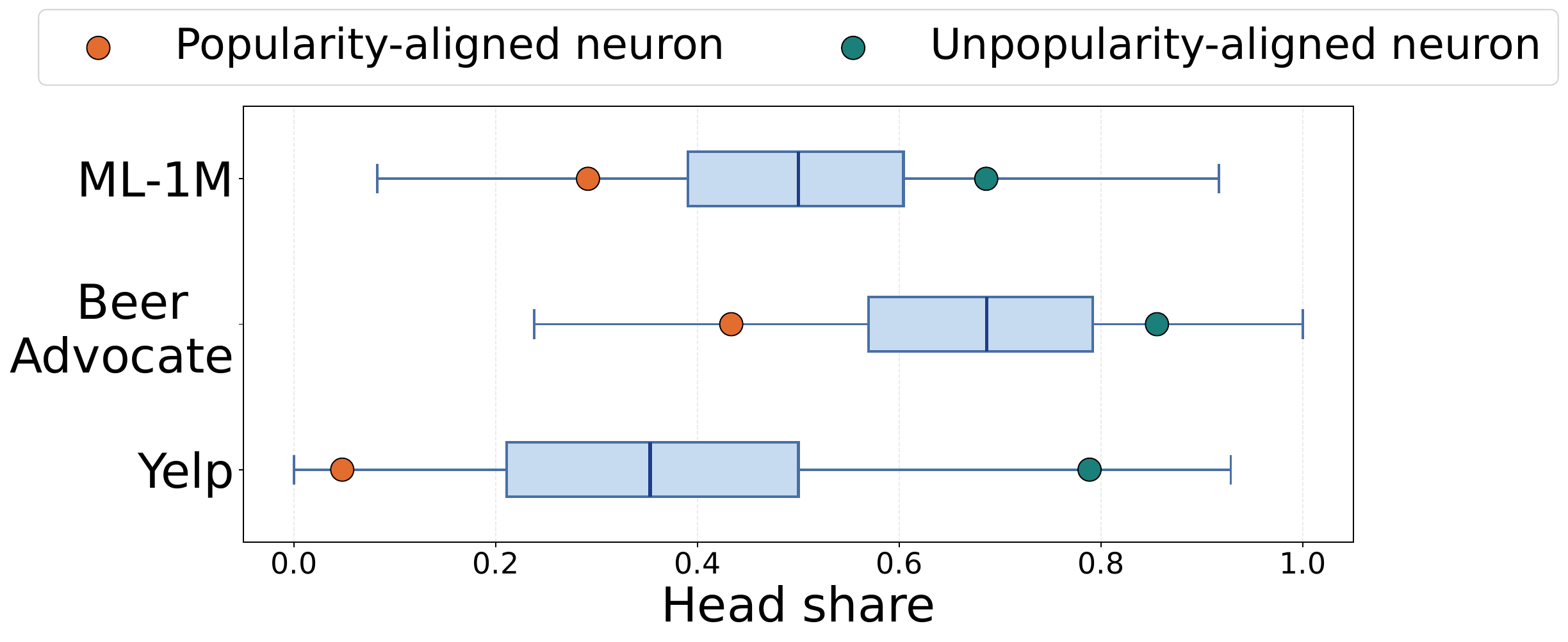}
    \caption{Head-item share $H$ over users for each dataset. Circles mark the mean $H$ of the top-10 users for the SAE neuron with the largest positive Cohen's $d$ (green, popularity-aligned) and the most negative Cohen's $d$ (orange, unpopularity-aligned).}
    \label{fig:box_plots}
\end{figure}

\begin{figure}[t]
    \centering
    \includegraphics[width=\linewidth]{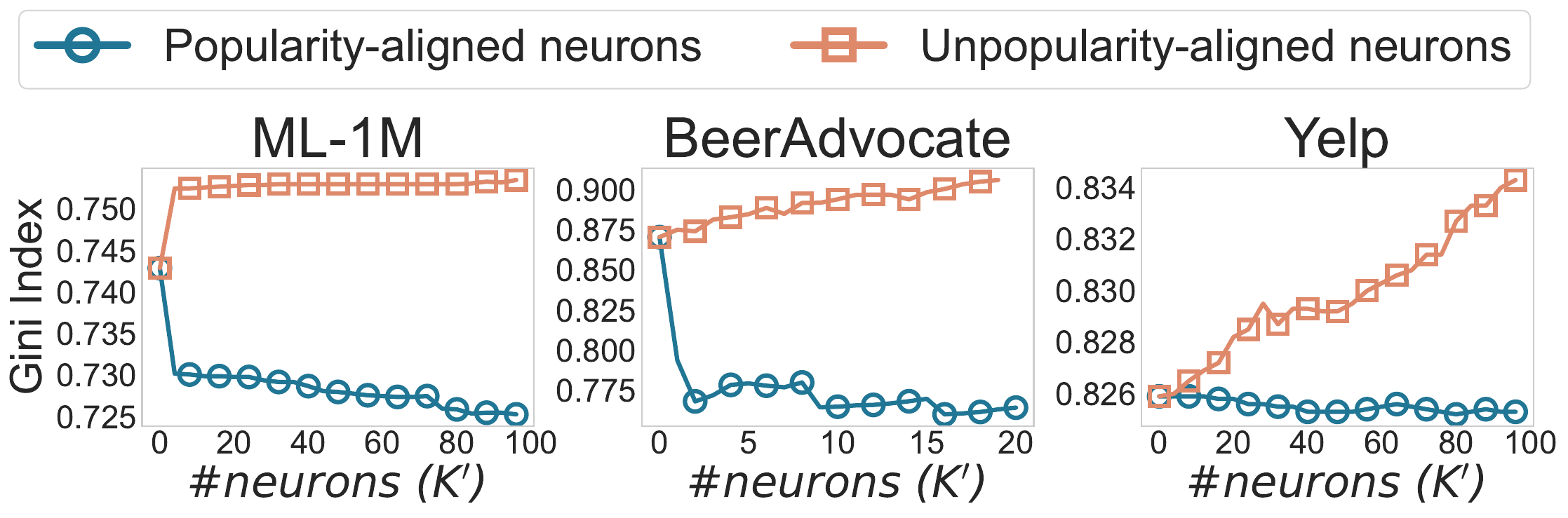}
    \caption{Effect of reducing activation of $K^\prime$ neurons linked to popularity bias ($\beta=1$).}
    \label{fig:interpretation}
\end{figure}

\begin{figure*}[t]
    \centering
    \vspace{-10pt}
    \includegraphics[width=.9\linewidth]{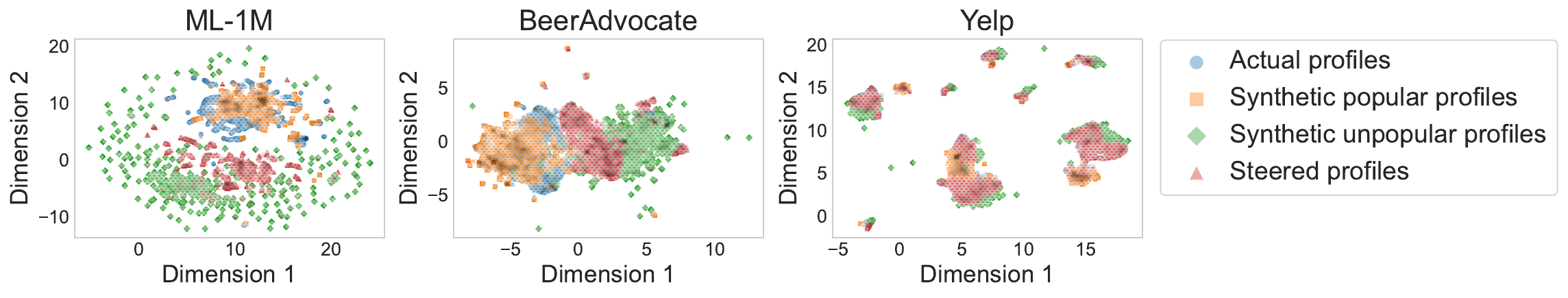}
        \caption{2D UMAP projections of SAE activations for real, synthetic popular/unpopular, and steered user profiles. For ML-1M and BeerAdvocate, $\alpha^{\text{Unpop}} = 3$, $\alpha^{\text{Pop}} = 3$. For Yelp, $\alpha^{\text{Unpop}} = 1$, $\alpha^{\text{Pop}} = 1$.}\label{fig:UMAP}
\end{figure*}

\subsection{Interpretability analysis}\label{sec:interpetability}
In this section, we address \textbf{RQ2}.

\subsubsection{Analyzing activations of targeted neurons} Each SAE neuron is assigned a Cohen’s $d$ value indicating its association with popular or unpopular items. To test whether these signals extend to real users, we select the neuron with the highest positive Cohen's $d$ and the one with the most negative Cohen's $d$. For all users in the validation set, we compute the head-item share $H$ (fraction of interactions with head items) and construct a boxplot over users, marking the mean $H$ of the top-10 most strongly activating users for each neuron. Intuitively, popularity-aligned neurons should fire more strongly for users who mostly consume head items, and unpopularity-aligned neurons should fire more strongly for users with tail-oriented histories.

Fig.~\ref{fig:box_plots} provides evidence that the identified neurons behave as intended on real users. For each dataset, the mean head share of the top-10 users for the most popularity-aligned neuron lies above the population median, often close to the upper whisker. In contrast, the mean head share for the most unpopularity-aligned neuron lies below the median and, for Yelp, near the lower extreme of the distribution. As expected, popularity-aligned neurons concentrate their highest activations on users with head-heavy histories, while unpopularity-aligned neurons focus on users who mainly interact with tail items.

\subsubsection{Manipulating activations of targeted neurons} Results from the previous subsection validates that we correctly identify biased neurons. To check whether the manipulation of selected neurons results in expected downstream performance, we conduct a controlled neuron manipulation study. Specifically, we manually reduce the activations of the top $K^\prime$ popularity-aligned ($d_j>1$) or unpopularity-aligned ($d_j<-1$) neurons by one standard deviation, i.e., $a_j^\prime=a_j-\sigma_j$, ensuring a bounded perturbation.

Figure \ref{fig:interpretation} illustrates how Gini Index evolves as the activation of these most influential neurons are progressively reduced. On all datasets, reducing the activations of popularity-aligned neurons consistently improves Gini Index, while reducing the activations of unpopularity-aligned neurons worsens Gini Index. Notably, ML-1M exhibits a sharp initial change, suggesting a small number of dominant neurons, which aligns with prior SAE findings on monosemantic representations \cite{bricken2023monosemanticity, gao2024scaling, o2024disentangling}.

\begin{table}[t]
  \caption{Wasserstein distance between data points in Fig.~\ref{fig:UMAP}: 0 indicates identical and higher value indicates more dissimilar distribution. Bolded entries show desirable values: higher distance to "Synthetic popular profiles" and lower distance to "Synthetic unpopular profiles".}
  \label{tab:umap_distance}
  \centering
  \small
  \setlength{\tabcolsep}{3pt}
  \begin{tabular}{l cccc}
    \toprule
     & ML-1M & BeerAdvocate & Yelp \\
    \midrule
    Actual → Synthetic popular & 1.53 & 1.61 & 2.11 \\
    Steered → Synthetic popular & \textbf{5.92} & \textbf{3.61} & \textbf{2.17} \\
    \hline
    Actual → Synthetic unpopular & 7.25 & 3.52 & 1.21 \\
    Steered → Synthetic unpopular & \textbf{3.42} & \textbf{1.53} & \textbf{1.16} \\
    \bottomrule
  \end{tabular}
\end{table}

\subsubsection{Analyzing user embedding space} We further analyze users' embedding space derived from SAE. Fig.~\ref{fig:UMAP} visualizes users' embedding using UMAP~\cite{mcinnes2018umap}, projecting real users, synthetic popular and unpopular profiles, and steered embeddings into 2D space.

Several patterns can be observed in Fig.~\ref{fig:UMAP}: 
\begin{enumerate*}[label=(\roman*)]
    \item synthetic unpopular and popular sequences form distinct clusters, confirming that our synthetic generation isolates popularity bias,
    \item actual profiles align closer to the popular cluster, reflecting inherent popularity bias, and
    \item after applying neuron steering with \algname{PopSteer}, the activations shift in the desired direction: they become more dissimilar to popular sequences and more similar to unpopular sequences.
\end{enumerate*}

This visual shift is quantitatively validated in Table \ref{tab:umap_distance}, which computes Wasserstein distances between distributions. These changes are statistically significant ($p<0.01$), validating the reliability of \algname{PopSteer} in mitigating bias. Overall, these findings demonstrate that \algname{PopSteer} not only mitigates popularity bias effectively but also provides actionable, interpretable insights into the neural basis of bias—enabling more transparent and controllable interventions.

\subsection{Ablation study}
To verify the effectiveness of our method, we conduct an ablation study with two variants to independently assess key components of \algname{PopSteer}: 
\begin{enumerate*}[label=(\roman*)]
    \item the neuron selection process (\algname{Noise}), and 
    \item the targeted steering logic (\algname{RandomSelect}). 
\end{enumerate*}

\begin{table}[t]
  \caption{Ablation study: \algname{PopSteer} (PS) vs.\@ \algname{Noise} vs.\@ \algname{RandomSelect} (RS).}
  \label{tab:pop_vs_noise}
  \centering
  \small
  \setlength{\tabcolsep}{4pt}
  \begin{tabular}{l r ccc ccc}
    \toprule
        &         & \multicolumn{3}{c}{Item Coverage~$\uparrow$} &
                      \multicolumn{3}{c}{Gini Index~$\downarrow$} \\
    \cmidrule(lr){3-5}\cmidrule(lr){6-8}
    Dataset & $ \beta $ &
    PS & Noise & RS &
    PS & Noise & RS \\
    \midrule
    \multirow{3}{*}{ML-1M}
      & 0.5 & \textbf{0.5707} & 0.4732 & 0.475 & \textbf{0.6645} & 0.7464 & 0.7429 \\
      & 1.0 & \textbf{0.5651} & 0.4741 & 0.4791 & \textbf{0.6698} & 0.7438 & 0.7442 \\
      & 1.5 & \textbf{0.5531} & 0.4750 & 0.475 & \textbf{0.6748} & 0.7428 & 0.7429 \\
    \midrule
    \multirow{3}{*}{BeerAdvocate}
      & 0.5 & \textbf{0.2998} & 0.2268 & 0.243 & \textbf{0.7846} & 0.8502 & 0.8409 \\
      & 1.0 & \textbf{0.2939} & 0.2131 & 0.2407 & \textbf{0.7910} & 0.8675 & 0.844 \\
      & 1.5 & \textbf{0.2849} & 0.2337 & 0.2315 & \textbf{0.7989} & 0.8433 & 0.8521 \\
    \midrule
    \multirow{3}{*}{Yelp}
      & 0.5 & \textbf{0.3582} & 0.3392 & 0.3377 & \textbf{0.8143} & 0.8273 & 0.8259 \\
      & 1.0 & \textbf{0.3589} & 0.3369 & 0.3398 & \textbf{0.8143} & 0.8268 & 0.8268 \\
      & 1.5 & \textbf{0.3406} & 0.3378 & 0.3449 & \textbf{0.8245} & 0.8261 & 0.8232 \\
    \bottomrule
  \end{tabular}
\end{table}

The first variant, denoted as \algname{Noise}, isolates the impact of our controlled steering mechanism as detailed in Eq.~\ref{steering_eq}. Here, we select the same set of neurons as in \algname{PopSteer} (based on highest absolute Cohen's $d$), but instead of applying bias-aware, directionally consistent modifications to their activations, 
we inject unstructured perturbations in the form of random Gaussian noise. We tune the hyperparameters of \algname{PopSteer} ($\alpha^{\text{Unpop}}$ and $\alpha^{\text{Pop}}$) and the Standard Deviation $\xi$ for the Noise variant such that the drop in nDCG@10 does not exceed 10\% relative to the base SASRec model.

The second variant, \algname{RandomSelect}, ablates the neuron selection logic by randomly sampling the same number of neurons as would be chosen under each $\beta$ threshold in \algname{PopSteer}. For these randomly selected neurons, it applies steering adjustments using the same weights $w_j$ and directions (suppress if $d_j > 0$, boost if $d_j < 0$) as those of the top neurons. This preserves the exact magnitudes, signs, and variance-normalized scales of the adjustments but applies them to arbitrary neurons instead, testing whether identifying the correct bias-encoding neurons is crucial for effectiveness.

Results are reported in Table \ref{tab:pop_vs_noise}, where we compare performance across different Cohen's $d$ thresholds ($\beta$) in terms of Item Coverage and Gini Index.
We observe that \algname{PopSteer} consistently yields superior fairness outcomes relative to baselines, accross all datasets. 
\algname{Noise} and \algname{RandomSelect} both perform similarly poorly, indicating that neither steering alone nor selection alone suffices. Consistent fairness gains emerge only when targeted steering is combined with Cohen’s $d$–guided neuron selection, as in \algname{PopSteer}.

\begin{figure}[t]
    \centering
    \begin{subfigure}[b]{.5\textwidth}
        \centering
        \includegraphics[width=\linewidth]{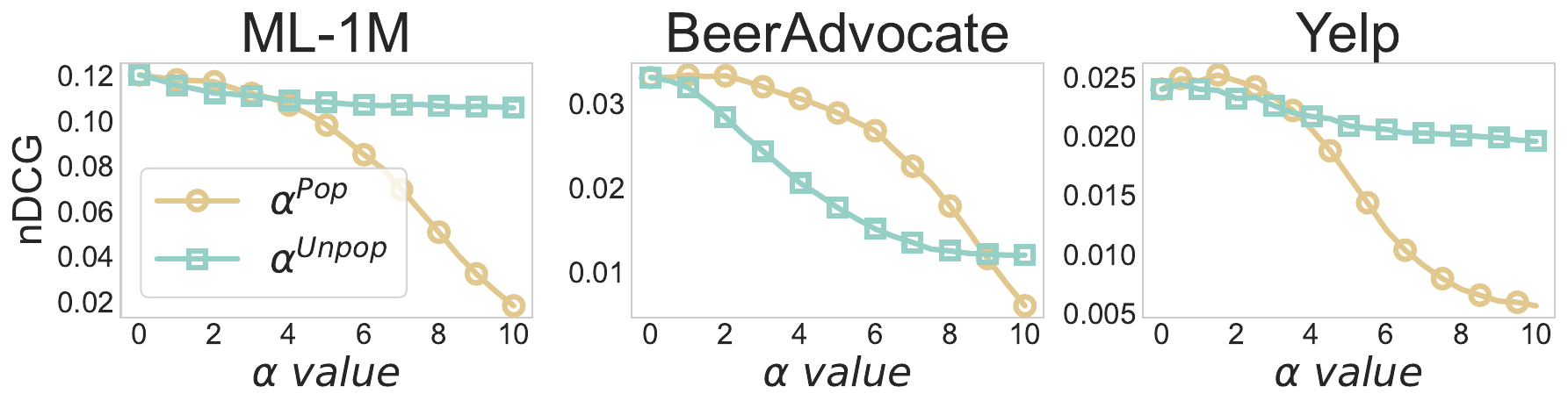}
        \vspace{-17pt}
        \caption{The effect of varying $\alpha^{\text{Pop}}$ and $\alpha^{\text{Unpop}}$ on nDCG.}\label{fig:alpha_effect_ndcg}
    \end{subfigure}
    \begin{subfigure}[b]{.5\textwidth}
        \centering
        \includegraphics[width=\linewidth]{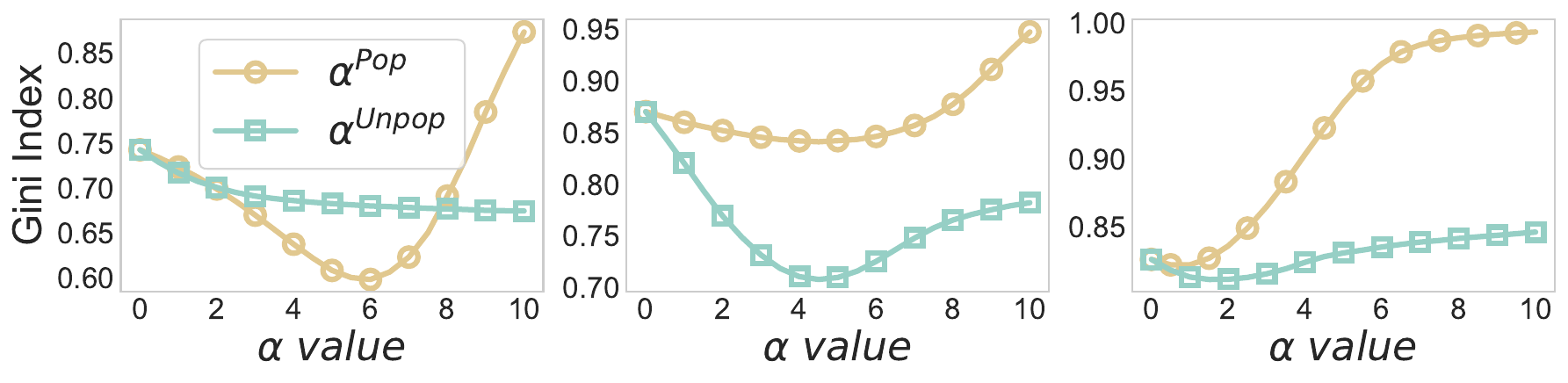}
        \vspace{-17pt}
        \caption{The effect of varying $\alpha^{\text{Pop}}$ and $\alpha^{\text{Unpop}}$ on Gini Index.}\label{fig:alpha_effect_gini}
    \end{subfigure}
    \caption{Performance analysis of \algname{PopSteer} in terms of nDCG and Gini Index with varying $\alpha^{\text{Pop}}$ and $\alpha^{\text{Unpop}}$.}\label{fig:alpha_effect}
\end{figure}

\subsection{Sensitivity analysis}\label{sensitivity_analysis}

This section addresses \textbf{RQ3}. Fig.~\ref{fig:alpha_effect} displays sensitivity analysis of $\alpha^{\text{Pop}}$ and $\alpha^{\text{Unpop}}$ across three datasets. Fig.~\ref{fig:alpha_effect_ndcg} shows the impact of these hyperparameters on nDCG, while Fig.~\ref{fig:alpha_effect_gini} illustrates effects on Gini Index. As these parameters increase, the Gini Index decreases, with predictable patterns. Reductions remain steady until dataset-specific thresholds, after which point gains plateau or reverse. 

The plots reveal an asymmetry: nDCG declines earlier and more sharply under larger $\alpha^{\text{Pop}}$—which suppresses popularity-aligned neurons—than under larger $\alpha^{\text{Unpop}}$, which amplifies long-tail neurons. This aligns with the model’s reliance on head-item features: excessive down-weighting harms relevance.

\section{Related Work}

\subsection{Interpretable recommender systems}
Interpretability in recommender systems seeks to expose the reasoning behind a model's predictions, enabling better trust, transparency, and actionable debugging~\cite{zhang2020explainable,liu2020explainable}.  
Approaches broadly fall into \textit{post-hoc} methods, which explain an existing black-box model's outputs without altering its architecture~\cite{zhang2019interpretable, jiang2023rcenr}, and \textit{intrinsically interpretable} models, whose structure is designed to be directly understandable~\cite{wang2018explainable,chen2019air}.  

Sparse Autoencoders, originally introduced for learning more compact and disentangled representations~\cite{ng2011sparse}, have recently emerged as a promising tool for interpretability. By enforcing strict sparsity in the hidden layer—often via top-$K$ activations functions~\cite{gao2024scaling}—SAEs encourage individual neurons to specialize in distinct, semantically coherent features. This ``monosemantic'' behavior has been leveraged in natural language processing and vision models to attribute concepts to specific neurons \cite{bricken2023monosemanticity, o2024disentangling, lan2024sparse}. Crucially, monosemanticity makes these neurons easier to isolate and manipulate without unintended side effects, a property already extensively leveraged to mitigate bias in language models ~\cite{zhao2024steering, kang2024interpret, durmus2024steering, harle2024scar}. 
While SAEs have seen increasing use in general-purpose model interpretability, their application to recommender systems remains underexplored, especially for diagnosing and intervening on systemic biases.

\subsection{Fairness in recommender systems}

Bias in recommender systems can arise from various sources, including historical user-item interactions, feedback loop, and algorithmic design~\cite{chen2023bias,yang2023rectifying,mansoury2020feedback}. A particularly well-studied form is \textit{popularity bias}, where popular items receive disproportionately high exposure, reducing recommendation diversity and limiting discovery of less-known content~\cite{abdollahpouri2021user}. This bias can negatively affect content providers and create echo-chamber effects for users~\cite{ge2020understanding,cinus2022effect}.

Mitigation strategies include re-ranking approaches that adjust recommendation lists to balance accuracy and fairness~\cite{zehlike2017fa,heuss2023predictive}, regularization techniques that penalize biased scoring function~\cite{abdollahpouri2017controlling}, and adversarial training to counteract biased signals~\cite{wu2021fight}. Some works optimize explicit fairness metrics such as exposure disparity~\cite{yadav2021policy,mansoury2024mitigating} or long-tail coverage~\cite{luo2023improving}. However, most existing methods treat the model as a black box, focusing on \emph{what} to change rather than \emph{why} the bias occurs. This leaves a gap for interpretability-driven interventions that can identify the internal mechanisms contributing to bias and steer them in a targeted manner.

\section{Conclusion and Future work}

This work introduces \algname{PopSteer}, a post-hoc method for interpreting and mitigating popularity bias in recommendation models using Sparse Autoencoder. By identifying and adjusting neuron activations linked to popularity signals, \algname{PopSteer} improves exposure fairness without sacrificing accuracy. Our experiments on public datasets demonstrate its effg
ectiveness and reliability compared to existing baselines. Beyond performance, the method offers interpretability and fine-grained control, making it a practical and transparent solution for fairness-aware recommendation.

Our synthetic user generation method isolates bias by removing confounding variables. While effective in our experiments, it could be refined using more sophisticated generation strategies. 
Future research could explore “smarter” generation, such as probabilistic sampling from item popularity gradients~\cite{ma2024negative} or generative models (e.g., GANs)~\cite{bobadilla2023creating} to produce distribution-aligned yet bias-amplified sequences.
Additionally, applying this framework for addressing other biases can be an interesting future research. For example, gender bias in recommendations could be mitigated by constructing synthetic datasets as described here to identify and steer bias-related neurons. We see this as a promising and impactful direction.

\bibliographystyle{ACM-Reference-Format}
\balance
\bibliography{ref}

\appendix

\input{fig_baselines_debias}

\section{Debiasing Effect of \algname{PopSteer}}\label{sec_debiasing}

According to~\cite{zhang2023empowering}, we additionally measure nDCG separately on Head and Tail items to evaluate the debiasing effect of our \algname{PopSteer} and baselines (see Section~\ref{sec_synthetic} for the definition of Head and Tail items).

Fig.~\ref{fig:baseline_debias} compares the overall nDCG with nDCG on head (Fig.~\ref{fig:baseline_debias-head}) and tail (Fig.~\ref{fig:baseline_debias_tail}) items, respectively. While debiasing recommendation models is not the primary focus of \algname{PopSteer}, it is interesting to see its potential in debiasing recommendation system. Fig.~\ref{fig:baseline_debias} reveals that \algname{PopSteer} enhances the accuracy of tail item recommendations, at a manageable drop in accuracy of head items recommendation. This shows that popularity bias mitigation by \algname{PopSteer} is not through merely shifting exposure, but through recommending tail items that are relevant and aligned with individual user preferences.

\section{Scalability analysis}\label{scalability_analysis}
To verify the scalability of \algname{PopSteer}, we compare its inference time to baselines. As shown in Fig.~\ref{fig:runtime}, \algname{PopSteer} achieves the lowest inference time compared to baselines (except \algname{Random}).  
Furthermore, for practicality, we also compute the training time of \algname{PopSteer} (in hours): 0.71 on ML-1M, 0.57 on BeerAdvocate, and 0.48 on Yelp. Despite the high scale factor $s$, training time remains manageable. This aligns with findings in~\cite{gao2024scaling}, which found that good initialization, top $K$ sparsity, and dead-latent prevention make large SAEs learn efficiently. 
Overall, both the training and inference times of \algname{PopSteer} are well within practical limits. This makes the use of \algname{PopSteer} highly suitable for production deployment.

\input{fig_runtime}

\section{SASRec hyperparameters}\label{sasrec_details}
Table \ref{tab:sasrec} contains the hyperparameters used for training SASRec.
\begin{table}[H]
\centering
\caption{SASRec Hyperparameters}
\label{tab:sasrec}
\begin{tabular}{ll}
\toprule
\textbf{Hyperparameter} & \textbf{Value} \\
\midrule
Number of transformer layers   & $2$ \\
Number of attention heads      & $2$ \\
Hidden size                    & $64$ \\
Feed-forward inner size        & $256$ \\
Hidden dropout probability     & $0.5$ \\
Attention dropout probability  & $0.5$ \\
Activation function            & \texttt{gelu} \\
Layer normalization $\epsilon$ & $1 \times 10^{-12}$ \\
Initializer standard deviation & $0.02$ \\
Loss function                  & Cross-Entropy \\
\bottomrule
\end{tabular}
\end{table}

\end{document}

%% file: fig_baselines_gini_.tex
\begin{figure*}[htbp]
    \begin{subfigure}[b]{1\textwidth}
        \centering

        \begin{tikzpicture}    
            \begin{axis}[
                hide axis,
                xmin=0, xmax=1, ymin=0, ymax=1,
                legend to name=mylegend,
                legend columns=8,
                legend style={
                    font=\normalsize,
                    text=black,  
                    column sep=0em,
                    row sep=0.6em,
                    draw=black,
                    fill=white,
                    legend cell align=left,
                    align = left,
                    nodes={inner sep=3pt},
                    text width=1.3cm,
                    at={(0.5,1.05)},
                    anchor=south,
                },
            ]
            \addlegendimage{only marks, mark=x, color={rgb, 255:red, 0; green, 128; blue, 128},mark size=3,line width=1.1pt}
            \addlegendentry{SASRec}
            \addlegendimage{only marks, mark=star, color={rgb, 255:red, 255; green, 182; blue, 193},mark size=3,line width=1.1pt}
            \addlegendentry{Random}
            \addlegendimage{only marks, mark=diamond*, fill=purple!40!white,mark size=3}
            \addlegendentry{PCT}
            \addlegendimage{only marks, mark=+, color={rgb, 75:red, 0; green, 130; blue, 128},mark size=2.5,line width=1.1pt}
            \addlegendentry{P-MMF}
            \addlegendimage{only marks, mark=square*, fill=yellow!40!white,mark size=2.5}
            \addlegendentry{IPR}
            \addlegendimage{only marks, mark=*, fill=orange!40!white,mark size=2.5}
            \addlegendentry{FA*IR}
            \addlegendimage{only marks, mark=triangle*, fill=green!40!white,mark size=3}
            \addlegendentry{DUOR}
            \addlegendimage{only marks, mark=square*, fill=blue!40!white,mark size=2.5}
            \addlegendentry{PopSteer}
            \end{axis}
        \end{tikzpicture}
    
        \ref{mylegend}
        \vspace{0.1em}
        
        \begin{tikzpicture}[scale=0.78]
            \begin{axis}[
                title={\bfseries\fontsize{11}{20}\selectfont ML-1M},
                xlabel=nDCG,
                ylabel=Item Coverage,
                width=6.2cm,
                height=4.7cm,
                legend pos=north west,
                legend cell align={left},
                legend style={fill opacity=0.75, draw opacity=1, text opacity=1},
                x label style={yshift=0.5em},
                y label style={yshift=-0.5em},
                scaled ticks=false,
                x tick label style={/pgf/number format/fixed, /pgf/number format/precision=3},
                y tick label style={/pgf/number format/fixed, /pgf/number format/precision=3},
            ]
            \addplot[
                scatter, only marks, scatter src=explicit symbolic,
                scatter/classes={
                    SASRec={mark=x, color={rgb, 255:red, 0; green, 128; blue, 128},mark size=3,line width=1.1pt},
                    Random={mark=star, color={rgb, 255:red, 255; green, 182; blue, 193},mark size=2.5,line width=1.1pt},
                    PCT={mark=diamond*, fill=purple!40!white,mark size=2.5},
                    PMMF={mark=+, color={rgb, 75:red, 0; green, 130; blue, 128},mark size=2.5,line width=1.1pt},
                    IPR={mark=square*, fill=yellow!40!white},
                    FA*IR={mark=*, fill=orange!40!white},
                    DUOR={mark=triangle*, fill=green!40!white,mark size=2.5},
                    PopSteer={mark=square*, fill=blue!40!white}
                }
            ]
            table [x=ndcg, y=cov5, col sep=comma, meta=approach] {baselines_ml.csv};
            \end{axis}
        \end{tikzpicture}
        \hspace{.2cm}%
        \begin{tikzpicture}[scale=0.78]
            \begin{axis}[
                title={\bfseries\fontsize{11}{20}\selectfont BeerAdvocate},
                xlabel=nDCG,
                width=6.2cm,
                height=4.7cm,
                legend pos=north west,
                legend cell align={left},
                legend style={fill opacity=0.75, draw opacity=1, text opacity=1},
                x label style={yshift=0.5em},
                y label style={yshift=-0.5em},
                scaled ticks=false,
                x tick label style={/pgf/number format/fixed, /pgf/number format/precision=3},
                y tick label style={/pgf/number format/fixed, /pgf/number format/precision=3},
            ]
            \addplot[
                scatter, only marks, scatter src=explicit symbolic,
                scatter/classes={
                    SASRec={mark=x, color={rgb, 255:red, 0; green, 128; blue, 128},mark size=3,line width=1.1pt},
                    Random={mark=star, color={rgb, 255:red, 255; green, 182; blue, 193},mark size=2.5,line width=1.1pt},
                    PCT={mark=diamond*, fill=purple!40!white,mark size=2.5},
                    PMMF={mark=+, color={rgb, 75:red, 0; green, 130; blue, 128},mark size=2.5,line width=1.1pt},
                    IPR={mark=square*, fill=yellow!40!white},
                    FA*IR={mark=*, fill=orange!40!white},
                    DUOR={mark=triangle*, fill=green!40!white,mark size=2.5},
                    PopSteer={mark=square*, fill=blue!40!white}
                }
            ]
            table [x=ndcg, y=cov5, col sep=comma, meta=approach] {baselines_beer.csv};
            \end{axis}
        \end{tikzpicture}
        \hspace{.2cm}%
        \begin{tikzpicture}[scale=0.78]
            \begin{axis}[
                title={\bfseries\fontsize{11}{20}\selectfont Yelp},
                xlabel=nDCG,
                width=6.2cm,
                height=4.7cm,
                legend pos=north west,
                legend cell align={left},
                legend style={fill opacity=0.75, draw opacity=1, text opacity=1},
                x label style={yshift=0.5em},
                y label style={yshift=-0.5em},
                scaled ticks=false,
                x tick label style={/pgf/number format/fixed, /pgf/number format/precision=3},
                y tick label style={/pgf/number format/fixed, /pgf/number format/precision=3},
            ]
            \addplot[
                scatter, only marks, scatter src=explicit symbolic,
                scatter/classes={
                    SASRec={mark=x, color={rgb, 255:red, 0; green, 128; blue, 128},mark size=3,line width=1.1pt},
                    Random={mark=star, color={rgb, 255:red, 255; green, 182; blue, 193},mark size=2.5,line width=1.1pt},
                    PCT={mark=diamond*, fill=purple!40!white,mark size=2.5},
                    PMMF={mark=+, color={rgb, 75:red, 0; green, 130; blue, 128},mark size=2.5,line width=1.1pt},
                    IPR={mark=square*, fill=yellow!40!white},
                    FA*IR={mark=*, fill=orange!40!white},
                    DUOR={mark=triangle*, fill=green!40!white,mark size=2.5},
                    PopSteer={mark=square*, fill=blue!40!white}
                }
            ]
            table [x=ndcg, y=cov5, col sep=comma, meta=approach] {baselines_yelp.csv};
            \end{axis}
        \end{tikzpicture}
        \vspace{-5pt}
        \caption{nDCG versus Item Coverage}
        \label{fig:baseline_cov}
    \end{subfigure}
    
    \begin{subfigure}[b]{1\textwidth}
        \centering        

        \begin{tikzpicture}[scale=0.78]
            \begin{axis}[
                xlabel=nDCG,
                ylabel=Gini Index,
                width=6.2cm,
                height=4.7cm,
                legend pos=north west,
                legend cell align={left},
                legend style={fill opacity=0.75, draw opacity=1, text opacity=1},
                x label style={yshift=0.5em},
                y label style={yshift=-0.5em},
                scaled ticks=false,
                x tick label style={/pgf/number format/fixed, /pgf/number format/precision=3},
                y tick label style={/pgf/number format/fixed, /pgf/number format/precision=3},
                xmin=0.06, xmax=0.13,
            ]
            \addplot[
                scatter, only marks, scatter src=explicit symbolic,
                scatter/classes={
                    SASRec={mark=x, color={rgb, 255:red, 0; green, 128; blue, 128},mark size=3,line width=1.1pt},
                    Random={mark=star, color={rgb, 255:red, 255; green, 182; blue, 193},mark size=2.5,line width=1.1pt},
                    PCT={mark=diamond*, fill=purple!40!white,mark size=2.5},
                    PMMF={mark=+, color={rgb, 75:red, 0; green, 130; blue, 128},mark size=2.5,line width=1.1pt},
                    IPR={mark=square*, fill=yellow!40!white},
                    FA*IR={mark=*, fill=orange!40!white},
                    DUOR={mark=triangle*, fill=green!40!white,mark size=2.5},
                    PopSteer={mark=square*, fill=blue!40!white}
                }
            ]
            table [x=ndcg, y=gini, col sep=comma, meta=approach] {baselines_ml.csv};
            \end{axis}
        \end{tikzpicture}
        \hspace{.2cm}%
        \begin{tikzpicture}[scale=0.78]
            \begin{axis}[
                xlabel=nDCG,
                width=6.2cm,
                height=4.7cm,
                legend pos=north west,
                legend cell align={left},
                legend style={fill opacity=0.75, draw opacity=1, text opacity=1},
                x label style={yshift=0.5em},
                y label style={yshift=-0.5em},
                scaled ticks=false,
                x tick label style={/pgf/number format/fixed, /pgf/number format/precision=3},
                y tick label style={/pgf/number format/fixed, /pgf/number format/precision=3},
                xmin=0.019, xmax=0.035,
                xtick={0.02,0.025,...,0.03},
            ]
            \addplot[
                scatter, only marks, scatter src=explicit symbolic,
                scatter/classes={
                    SASRec={mark=x, color={rgb, 255:red, 0; green, 128; blue, 128},mark size=3,line width=1.1pt},
                    Random={mark=star, color={rgb, 255:red, 255; green, 182; blue, 193},mark size=2.5,line width=1.1pt},
                    PCT={mark=diamond*, fill=purple!40!white,mark size=2.5},
                    PMMF={mark=+, color={rgb, 75:red, 0; green, 130; blue, 128},mark size=2.5,line width=1.1pt},
                    IPR={mark=square*, fill=yellow!40!white},
                    FA*IR={mark=*, fill=orange!40!white},
                    DUOR={mark=triangle*, fill=green!40!white,mark size=2.5},
                    PopSteer={mark=square*, fill=blue!40!white}
                }
            ]
            table [x=ndcg, y=gini, col sep=comma, meta=approach] {baselines_beer.csv};
            \end{axis}
        \end{tikzpicture}
        \hspace{.2cm}%
        \begin{tikzpicture}[scale=0.78]
            \begin{axis}[
                xlabel=nDCG,
                width=6.2cm,
                height=4.7cm,
                legend pos=north west,
                legend cell align={left},
                legend style={fill opacity=0.75, draw opacity=1, text opacity=1},
                x label style={yshift=0.5em},
                y label style={yshift=-0.5em},
                scaled ticks=false,
                x tick label style={/pgf/number format/fixed, /pgf/number format/precision=3},
                y tick label style={/pgf/number format/fixed, /pgf/number format/precision=3},
                xmin=0.016, xmax=0.0255,
                xtick={0.016,0.018,...,0.024},
            ]
            \addplot[
                scatter, only marks, scatter src=explicit symbolic,
                scatter/classes={
                    SASRec={mark=x, color={rgb, 255:red, 0; green, 128; blue, 128},mark size=3,line width=1.1pt},
                    Random={mark=star, color={rgb, 255:red, 255; green, 182; blue, 193},mark size=2.5,line width=1.1pt},
                    PCT={mark=diamond*, fill=purple!40!white,mark size=2.5},
                    PMMF={mark=+, color={rgb, 75:red, 0; green, 130; blue, 128},mark size=2.5,line width=1.1pt},
                    IPR={mark=square*, fill=yellow!40!white},
                    FA*IR={mark=*, fill=orange!40!white},
                    DUOR={mark=triangle*, fill=green!40!white,mark size=2.5},
                    PopSteer={mark=square*, fill=blue!40!white}
                }
            ]
            table [x=ndcg, y=gini, col sep=comma, meta=approach] {baselines_yelp.csv};
            \end{axis}
        \end{tikzpicture}
        \vspace{-5pt}
        \caption{nDCG versus Gini Index}
        \label{fig:baseline_gini}
    \end{subfigure}

\caption{Performance comparison of \algname{PopSteer} with baselines, in terms of nDCG versus (a) item coverage and (b) Gini Index.}
\label{fig:baseline_tradeoffs}
\end{figure*}

%% file: fig_baselines_debias.tex
\begin{figure*}[btp]
    \begin{subfigure}[b]{1\textwidth}
        \centering

        \begin{tikzpicture}    
            \begin{axis}[
                hide axis,
                xmin=0, xmax=1, ymin=0, ymax=1,
                legend to name=mylegend,
                legend columns=8,
                legend style={
                    font=\normalsize,
                    text=black,  
                    column sep=0em,
                    row sep=0.6em,  
                    draw=black,        
                    fill=white,        
                    legend cell align=left,
                    align = left,
                    nodes={inner sep=3pt}, 
                    text width=1.7cm,    
                    at={(0.5,1.05)},  
                    anchor=south,
                },
            ]
            \addlegendimage{only marks, mark=x, color={rgb, 255:red, 0; green, 128; blue, 128},mark size=3,line width=1.1pt}
            \addlegendentry{SASRec}
            \addlegendimage{only marks, mark=star, color={rgb, 255:red, 255; green, 182; blue, 193},mark size=3,line width=1.1pt}
            \addlegendentry{Random}
            \addlegendimage{only marks, mark=diamond*, fill=purple!40!white,mark size=3}
            \addlegendentry{PCT}
            \addlegendimage{only marks, mark=+, color={rgb, 75:red, 0; green, 130; blue, 128},mark size=2.5,line width=1.1pt}
            \addlegendentry{P-MMF}
            \addlegendimage{only marks, mark=square*, fill=yellow!40!white,mark size=2.5}
            \addlegendentry{IPR}
            \addlegendimage{only marks, mark=*, fill=orange!40!white,mark size=2.5}
            \addlegendentry{FA*IR}
            \addlegendimage{only marks, mark=triangle*, fill=green!40!white,mark size=3}
            \addlegendentry{DUOR}
            \addlegendimage{only marks, mark=square*, fill=blue!40!white,mark size=2.5}
            \addlegendentry{PopSteer}
            \end{axis}
        \end{tikzpicture}
    
        \ref{mylegend}
        \vspace{0.1em}
        
        \begin{tikzpicture}[scale=1]
            \begin{axis}[
                title={\bfseries\fontsize{11}{20}\selectfont ML-1M},
                xlabel=nDCG,
                ylabel=nDCG Head,
                width=6cm,
                height=4.5cm,
                legend pos=north west,
                legend cell align={left},
                legend style={fill opacity=0.75, draw opacity=1, text opacity=1},
                x label style={yshift=0.5em},
                y label style={yshift=-0.5em},
                scaled ticks=false,
                x tick label style={/pgf/number format/fixed, /pgf/number format/precision=3},
                y tick label style={/pgf/number format/fixed, /pgf/number format/precision=3},
            ]
            \addplot[
                scatter, only marks, scatter src=explicit symbolic,
                scatter/classes={
                    SASRec={mark=x, color={rgb, 255:red, 0; green, 128; blue, 128},mark size=3,line width=1.1pt},
                    Random={mark=star, color={rgb, 255:red, 255; green, 182; blue, 193},mark size=2.5,line width=1.1pt},
                    PCT={mark=diamond*, fill=purple!40!white,mark size=2.5},
                    PMMF={mark=+, color={rgb, 75:red, 0; green, 130; blue, 128},mark size=2.5,line width=1.1pt},
                    IPR={mark=square*, fill=yellow!40!white},
                    FA*IR={mark=*, fill=orange!40!white},
                    DUOR={mark=triangle*, fill=green!40!white,mark size=2.5},
                    PopSteer={mark=square*, fill=blue!40!white}
                }
            ]
            table [x=ndcg, y=ndcghead, col sep=comma, meta=approach] {baselines_ml_debias.csv};
            \end{axis}
        \end{tikzpicture}
        \hspace{.2cm}
        \begin{tikzpicture}[scale=1]
            \begin{axis}[
                title={\bfseries\fontsize{11}{20}\selectfont BeerAdvocate},
                xlabel=nDCG,
                width=6cm,
                height=4.5cm,
                legend pos=north west,
                legend cell align={left},
                legend style={fill opacity=0.75, draw opacity=1, text opacity=1},
                x label style={yshift=0.5em},
                y label style={yshift=-0.5em},
                scaled ticks=false,
                x tick label style={/pgf/number format/fixed, /pgf/number format/precision=3},
                y tick label style={/pgf/number format/fixed, /pgf/number format/precision=3},
            ]
            \addplot[
                scatter, only marks, scatter src=explicit symbolic,
                scatter/classes={
                    SASRec={mark=x, color={rgb, 255:red, 0; green, 128; blue, 128},mark size=3,line width=1.1pt},
                    Random={mark=star, color={rgb, 255:red, 255; green, 182; blue, 193},mark size=2.5,line width=1.1pt},
                    PCT={mark=diamond*, fill=purple!40!white,mark size=2.5},
                    PMMF={mark=+, color={rgb, 75:red, 0; green, 130; blue, 128},mark size=2.5,line width=1.1pt},
                    IPR={mark=square*, fill=yellow!40!white},
                    FA*IR={mark=*, fill=orange!40!white},
                    DUOR={mark=triangle*, fill=green!40!white,mark size=2.5},
                    PopSteer={mark=square*, fill=blue!40!white}
                }
            ]
            table [x=ndcg, y=ndcghead, col sep=comma, meta=approach] {baselines_beer_debias.csv};
            \end{axis}
        \end{tikzpicture}
        \hspace{.2cm}
        \begin{tikzpicture}[scale=1]
            \begin{axis}[
                title={\bfseries\fontsize{11}{20}\selectfont Yelp},
                xlabel=nDCG,
                width=6cm,
                height=4.5cm,
                legend pos=north west,
                legend cell align={left},
                legend style={fill opacity=0.75, draw opacity=1, text opacity=1},
                x label style={yshift=0.5em},
                y label style={yshift=-0.5em},
                scaled ticks=false,
                x tick label style={/pgf/number format/fixed, /pgf/number format/precision=3},
                y tick label style={/pgf/number format/fixed, /pgf/number format/precision=3},
            ]
            \addplot[
                scatter, only marks, scatter src=explicit symbolic,
                scatter/classes={
                    SASRec={mark=x, color={rgb, 255:red, 0; green, 128; blue, 128},mark size=3,line width=1.1pt},
                    Random={mark=star, color={rgb, 255:red, 255; green, 182; blue, 193},mark size=2.5,line width=1.1pt},
                    PCT={mark=diamond*, fill=purple!40!white,mark size=2.5},
                    PMMF={mark=+, color={rgb, 75:red, 0; green, 130; blue, 128},mark size=2.5,line width=1.1pt},
                    IPR={mark=square*, fill=yellow!40!white},
                    FA*IR={mark=*, fill=orange!40!white},
                    DUOR={mark=triangle*, fill=green!40!white,mark size=2.5},
                    PopSteer={mark=square*, fill=blue!40!white}
                }
            ]
            table [x=ndcg, y=ndcghead, col sep=comma, meta=approach] {baselines_yelp.csv};
            \end{axis}
        \end{tikzpicture}
        \vspace{-5pt}
        \caption{nDCG versus nDCG Head}
        \label{fig:baseline_debias-head}
    \end{subfigure}
    
    \begin{subfigure}[b]{1\textwidth}
        \centering        
        \begin{tikzpicture}[scale=1]
            \begin{axis}[
                xlabel=nDCG,
                ylabel=nDCG Tail,
                width=6cm,
                height=4.5cm,
                legend pos=north west,
                legend cell align={left},
                legend style={fill opacity=0.75, draw opacity=1, text opacity=1},
                x label style={yshift=0.5em},
                y label style={yshift=-0.5em},
                scaled ticks=false,
                x tick label style={/pgf/number format/fixed, /pgf/number format/precision=3},
                y tick label style={/pgf/number format/fixed, /pgf/number format/precision=3},
            ]
            \addplot[
                scatter, only marks, scatter src=explicit symbolic,
                scatter/classes={
                    SASRec={mark=x, color={rgb, 255:red, 0; green, 128; blue, 128},mark size=3,line width=1.1pt},
                    Random={mark=star, color={rgb, 255:red, 255; green, 182; blue, 193},mark size=2.5,line width=1.1pt},
                    PCT={mark=diamond*, fill=purple!40!white,mark size=2.5},
                    PMMF={mark=+, color={rgb, 75:red, 0; green, 130; blue, 128},mark size=2.5,line width=1.1pt},
                    IPR={mark=square*, fill=yellow!40!white},
                    FA*IR={mark=*, fill=orange!40!white},
                    DUOR={mark=triangle*, fill=green!40!white,mark size=2.5},
                    PopSteer={mark=square*, fill=blue!40!white}
                }
            ]
            table [x=ndcg, y=ndcgtail, col sep=comma, meta=approach] {baselines_ml_debias.csv};
            \end{axis}
        \end{tikzpicture}
        \hspace{.2cm}
        \begin{tikzpicture}[scale=1]
            \begin{axis}[
                xlabel=nDCG,
                width=6cm,
                height=4.5cm,
                legend pos=north west,
                legend cell align={left},
                legend style={fill opacity=0.75, draw opacity=1, text opacity=1},
                x label style={yshift=0.5em},
                y label style={yshift=-0.5em},
                scaled ticks=false,
                x tick label style={/pgf/number format/fixed, /pgf/number format/precision=3},
                y tick label style={/pgf/number format/fixed, /pgf/number format/precision=3},
            ]
            \addplot[
                scatter, only marks, scatter src=explicit symbolic,
                scatter/classes={
                    SASRec={mark=x, color={rgb, 255:red, 0; green, 128; blue, 128},mark size=3,line width=1.1pt},
                    Random={mark=star, color={rgb, 255:red, 255; green, 182; blue, 193},mark size=2.5,line width=1.1pt},
                    PCT={mark=diamond*, fill=purple!40!white,mark size=2.5},
                    PMMF={mark=+, color={rgb, 75:red, 0; green, 130; blue, 128},mark size=2.5,line width=1.1pt},
                    IPR={mark=square*, fill=yellow!40!white},
                    FA*IR={mark=*, fill=orange!40!white},
                    DUOR={mark=triangle*, fill=green!40!white,mark size=2.5},
                    PopSteer={mark=square*, fill=blue!40!white}
                }
            ]
            table [x=ndcg, y=ndcgtail, col sep=comma, meta=approach] {baselines_beer_debias.csv};
            \end{axis}
        \end{tikzpicture}
        \hspace{.2cm}
        \begin{tikzpicture}[scale=1]
            \begin{axis}[
                xlabel=nDCG,
                width=6cm,
                height=4.5cm,
                legend pos=north west,
                legend cell align={left},
                legend style={fill opacity=0.75, draw opacity=1, text opacity=1},
                x label style={yshift=0.5em},
                y label style={yshift=-0.5em},
                scaled ticks=false,
                x tick label style={/pgf/number format/fixed, /pgf/number format/precision=3},
                y tick label style={/pgf/number format/fixed, /pgf/number format/precision=3},
            ]
            \addplot[
                scatter, only marks, scatter src=explicit symbolic,
                scatter/classes={
                    SASRec={mark=x, color={rgb, 255:red, 0; green, 128; blue, 128},mark size=3,line width=1.1pt},
                    Random={mark=star, color={rgb, 255:red, 255; green, 182; blue, 193},mark size=2.5,line width=1.1pt},
                    PCT={mark=diamond*, fill=purple!40!white,mark size=2.5},
                    PMMF={mark=+, color={rgb, 75:red, 0; green, 130; blue, 128},mark size=2.5,line width=1.1pt},
                    IPR={mark=square*, fill=yellow!40!white},
                    FA*IR={mark=*, fill=orange!40!white},
                    DUOR={mark=triangle*, fill=green!40!white,mark size=2.5},
                    PopSteer={mark=square*, fill=blue!40!white}
                }
            ]
            table [x=ndcg, y=ndcgtail, col sep=comma, meta=approach] {baselines_yelp_debias.csv};
            \end{axis}
        \end{tikzpicture}
        \vspace{-5pt}
        \caption{nDCG versus nDCG Tail}
        \label{fig:baseline_debias_tail}
    \end{subfigure}
\caption{Performance comparison of \algname{PopSteer} with baselines in terms of overall nDCG and nDCG on Head and Tail items.}
\label{fig:baseline_debias}
\end{figure*}
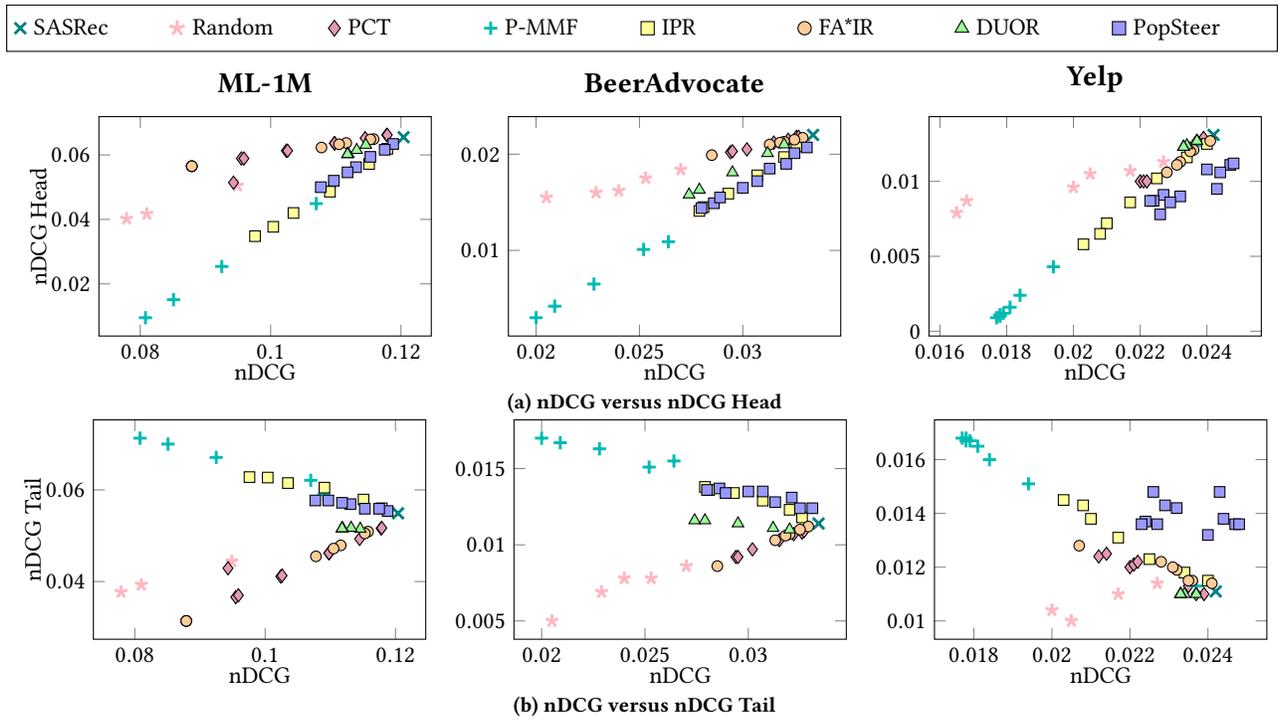

%% file: fig_runtime.tex
\begin{figure}[H]
    \centering

\begin{tikzpicture}
\begin{axis}[
    ymode=log,
    ymin=0.001, ymax=400,
    ylabel={\textcolor{gray}{Log} running time (seconds)},
    symbolic x coords={ML, Beer, Yelp},
    xtick=data,
    ymajorgrids=true,
    grid style={dashed,gray!50},
    width=9cm,
    height=4.5cm,
    bar width=5pt, 
    enlarge x limits=0.25,
    legend cell align=left,
    legend style={
        at={(0.5,1.05)},
        anchor=south,
        legend columns=4,
        font=\footnotesize,
        /tikz/every even column/.append style={column sep=6pt},
    },
    xticklabel style={font=\large},
    yticklabel style={font=\footnotesize},
]

\addplot[
    ybar,
    bar shift=-21pt,
    fill={rgb,255:red,0;green,128;blue,128},
    draw=black,
    thick,
    legend rect={sasrec!100!white}
] coordinates {(ML,0.0024) (Beer,0.0034) (Yelp,0.0063)};

\addplot[
    ybar,
    bar shift=-15pt,
    fill={rgb,255:red,255;green,182;blue,193},
    draw=black,
    thick,
    legend rect={random!100!white}
] coordinates {(ML,0.1531) (Beer,0.2945) (Yelp,0.3266)};

\addplot[
    ybar,
    bar shift=-9pt,
    fill=purple!40!white,
    draw=black,
    thick,
    legend rect={purple!40!white}
] coordinates {(ML,6.268) (Beer,19.929) (Yelp,33.873)};

\addplot[
    ybar,
    bar shift=-3pt,
    fill={rgb, 75:red, 0; green, 130; blue, 128},
    draw=black,
    thick,
    legend rect={duor!40!white}
] coordinates {(ML,16.764) (Beer,35.829) (Yelp,73.570)};

\addplot[
    ybar,
    bar shift=3pt,
    fill=yellow!40!white,
    draw=black,
    thick,
    legend rect={yellow!40!white}
] coordinates {(ML,0.420) (Beer,1.345) (Yelp,2.688)};

\addplot[
    ybar,
    bar shift=9pt,
    fill=orange!40!white,
    draw=black,
    thick,
    legend rect={orange!40!white}
] coordinates {(ML,6.853) (Beer,17.690) (Yelp,38.025)};

\addplot[
    ybar,
    bar shift=15pt,
    fill=green!40!white,
    draw=black,
    thick,
    legend rect={green!40!white}
] coordinates {(ML,9.462787628) (Beer,17.0398922) (Yelp,33.64384842)};

\addplot[
    ybar,
    bar shift=21pt,
    fill=blue!40!white,
    draw=black,
    thick,
    legend rect={blue!40!white}
] coordinates {(ML,0.405) (Beer,0.499) (Yelp,0.468)};

\legend{SASRec, Random, PCT, PMMF, IPR, FA*IR, DUOR, PopSteer}

\end{axis}
\end{tikzpicture}
\caption{Test set inference time of \algname{PopSteer} versus baselines.}
\label{fig:runtime}
\end{figure}
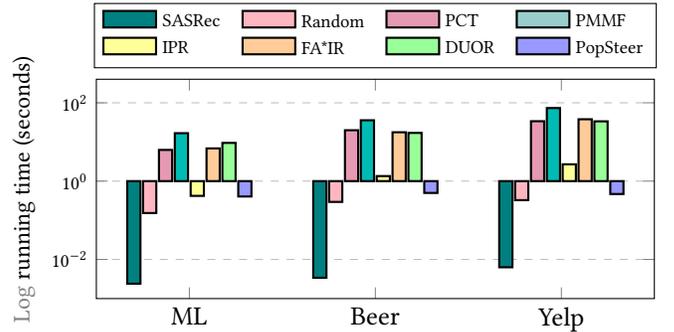